\documentclass[unnumsec,contemporary,large]{oup-authoring-template}%
\graphicspath{{Fig/}}

\theoremstyle{thmstyleone}%
\theoremstyle{thmstyletwo}%
\theoremstyle{thmstylethree}%

\begin{document}

\journaltitle{}
\DOI{DOI added during production}
\copyrightyear{YEAR}
\pubyear{YEAR}
\vol{XX}
\issue{x}
\access{Published: Date added during production}
\appnotes{Paper}

\firstpage{1}

%\subtitle{Subject Section}

\title[Causal Framework for Consciousness]{Avoiding Epiphenomenalism in Theories of Consciousness: A Causal Framework Based on Asymmetry}

\author[1,$\ast$]{Yoshiyuki Ohmura \ORCID{0000-0002-9158-5360}}
\author[1]{Yasuo Kuniyoshi}

\address[1]{\orgdiv{Department of Mechano Informatics}, \orgname{The University of Tokyo}, \orgaddress{\street{Hogo 7-3-1}, \postcode{113-8656}, \state{Tokyo}, \country{Japan}}}

\corresp[$\ast$]{Corresponding author. \href{email:ohmura@isi.imi.i.u-tokyo.ac.jp}{ohmura@isi.imi.i.u-tokyo.ac.jp}}

\received{Date}{0}{Year}
\revised{Date}{0}{Year}
\accepted{Date}{0}{Year}

%\editor{Associate Editor: Name}

%\abstract{
%\textbf{Motivation:} .\\
%\textbf{Results:} .\\
%\textbf{Availability:} .\\
%\textbf{Contact:} \href{name@email.com}{name@email.com}\\
%\textbf{Supplementary information:} Supplementary data are available at \textit{Journal Name}
%online.}

\abstract{
Why do some physical systems possess consciousness, while others do not? 
A scientific theory of consciousness must explain differences in system behavior in terms of well-defined internal mechanisms. 
However, many existing approaches introduce higher-level structural or informational descriptions without specifying how these structures contribute to the generation of behavior, raising the concern that they may become theoretically ineffective.
A further difficulty arises from the status of causation. 
Attempts to introduce causally effective mental states appear to conflict with physical theory, while strictly physicalist accounts risk excluding higher-level causation altogether. 
We argue that this dilemma stems from a conflation of physical determination with causation.
To address this issue, we distinguish between two complementary descriptive frameworks: the physical stance and the causal stance. 
While the physical stance represents systems in terms of state evolution, the causal stance introduces asymmetric relations that enable the explicit representation of internal mechanisms. 
We show that causal descriptions provide a more constrained and informative representation of internal structure and cannot, in general, be reconstructed from physical descriptions alone.
Within this framework, we define mental causation in terms of intrinsic causes and propose the Dual-Laws Model (DLM), which explicitly implements whole-to-parts causal mechanisms. 
This approach provides a principled way to model the causal role of consciousness without violating physical determinism.
} 


%\keywords[Abbreviations]{abbreviation1, abbreviation2, abbreviation3, abbreviation4}

%\otherabstract[Additional Abstract]{Use this element for elements such as Graphical abstract, Lay summary, Translated abstract etc. Que cum aut etum qui ium dolupta ssequia autati odis demporepe ad et es alit rem repudaerae min et volorum re volupta nobit volectur aut fuga.}

%\otherabstract[Graphical Abstract]{\colorbox{black!20}{\hbox to 0.97\textwidth{\vbox to 50pt{}}}}

\boxedtext{Highlights:}{
\begin{itemize}
\item We argue that physical theories do not define causal relations, but only state evolution.  
\item We identify a central limitation of existing approaches: internal structure often lacks causal efficacy and becomes theoretically ineffective.
\item We introduce the Causal Stance, which represents systems using asymmetric causal relations distinct from physical descriptions.
\item We argue that causal descriptions provide a more constrained and informative representation of internal mechanisms than dynamical models.
\item We propose a Dual-Laws Model (DLM) that implements intrinsic whole-to-parts causation within this framework. 
\end{itemize}}

\maketitle

%\begin{epigraph}
%Epigraph text. Ximporem qui reperov idempedit modio. Bisto imagnatem quae aceptis
%nobitae quid eum rae adignis quias-sit vellacc uptatur sunt quis rentis eaquasit alia deliquam
%rec-to consed unt. Empor sum ratur ressimusdae. Nam fugiae.
%\source{Epigraph source}
%\end{epigraph}

\section{Introduction}
Scientific theories are evaluated not only in terms of empirical testability, but also in terms of the role played by their theoretical constructs (\cite{woodward, giere}). 
Concepts that are poorly defined or that fail to contribute to explanatory structure have often been abandoned, not merely because of evidential difficulties, but because they ceased to play a productive role within scientific explanations (\cite{salmon, fraassen}).
These theoretical requirements concern the internal structure of a theory and can therefore be assessed independently of empirical testing. They provide a way of evaluating scientific theories even in domains where empirical assessment remains difficult.

In the study of consciousness, recent approaches have emphasized the importance of internal structure (\cite{tononi1998, hoel, kleiner2024, Ellia2025StructuralTurn, ellia2026, Tononi2025ConsciousnessOrPseudoConsciousness}). 
These frameworks incorporate informational, or organizational features of the system, moving beyond purely input–output characterizations. 
While they provide increasingly rich descriptions of internal organization (\cite{rosas, albantakis, Mediano2025}), differences in internal structure do not always correspond to differences in system behavior (\cite{Doerig2019, herzog2022, Klincewicz2025Unscientific}).
If such structures are independent of the mechanisms that generate system behavior (\cite{Tononi2025ConsciousnessOrPseudoConsciousness, ellia2026}), their explanatory significance remains unclear.

Before proposing a theory of consciousness, we must clarify what counts as a scientifically acceptable theory of consciousness (for a detailed discussion of theoretical epiphenomenalism and the criteria, see \cite{ohmura2026_te}).
This requirement can be assessed prior to empirical testability (\cite{kleiner,hoel2025}) \footnote{Theoretical effectiveness is closely related to empirical testability. If a theoretical construct makes no contribution to explanatory structure or predicted system behavior, it becomes unclear how that construct itself could be empirically evaluated.}. 
Whether a theory is empirically testable may depend on both the structure of the theory and the nature of the phenomenon under investigation. By contrast, the present requirement concerns the explanatory role of theoretical constructs within a scientific model, independently of the particular difficulties associated with consciousness research.

In many areas of science, theoretical constructs are expected to contribute to the explanatory structure of a model. 
Surprisingly, this requirement has rarely been examined explicitly in theories of consciousness. 
One reason may be that introducing causally effective consciousness-related constructs appears to conflict with widely held assumptions about physical determinism. 
As a result, consciousness-related concepts are often discussed in terms of characterization rather than explanatory contribution.

The goal of the present work is not to offer a definitive theory of consciousness, but to show that consciousness-related concepts can be incorporated into a scientific theory without becoming theoretically ineffective constructs.

Before proceeding, we wish to clarify the status of consciousness within the present framework \footnote{The purpose of the present framework is not to establish what mind or consciousness ultimately is. Rather, it is to clarify what kind of representational framework is required if consciousness-related concepts are to play a causal and explanatory role within a scientific theory.}. 
Scientific theories are representational frameworks that describe phenomena through theoretical entities and relations. 
Consequently, the concepts appearing in a theory should not be identified with the phenomena themselves. This observation is not unique to consciousness research. For example, the variable $m$ in physics is a theoretical representation of mass; it does not itself possess weight. 
Yet this does not imply that physical theories are unrelated to the phenomenon of mass. Similarly, the concepts introduced in a theory of consciousness should not be regarded as phenomenal experience itself. At the same time, this does not imply that they are unrelated to conscious experience or qualia. Rather, they are theoretical representations intended to characterize and explain phenomena associated with consciousness. We therefore focus on developing a causal framework capable of explaining the role of consciousness within a scientific model, while remaining agnostic regarding the status of phenomenal qualities considered in isolation.

\section{The Problem of Causation in Theories of Mind}
Discussions influenced by the hard problem of consciousness (\cite{chalmers1995}) have often emphasized qualia and phenomenal experience. 
While such concerns are important for understanding consciousness, a scientific theory must do more than merely characterize a phenomenon. Theoretical constructs introduced within a model should contribute to the explanation of how system behavior and internal organization are generated.
Mere association with conscious experience is not sufficient to justify their inclusion. Otherwise, such constructs risk becoming theoretically ineffective, describing aspects of consciousness without contributing to the explanatory structure of the theory.

The requirement that theoretical constructs must contribute to system behavior (\cite{woodward}) raises a difficulty when applied to theories of mind and consciousness. 
Any attempt to introduce causally effective higher-level constructs--such as mental states or system-level organization--must confront a well-known tension concerning the status of causation itself.

On the one hand, attributing causal efficacy to non-physical entities appears to commit one to a form of dualism. 
If mental states are treated as causes distinct from physical processes, then the framework seems to posit causal interactions beyond the scope of physical theory, thereby undermining its status as a scientific account.

On the other hand, adopting a strictly physicalist perspective leads to an opposing difficulty. 
If all physical processes are fully determined by physical states and laws, then any additional higher-level causes appear either redundant or superfluous. 
As emphasized by \cite{kim1998}, if a physical cause is sufficient to produce an effect, any higher-level cause either duplicates this role or fails to contribute independently.

This generates a dilemma. 
Introducing causally effective higher-level constructs appears to conflict with physical determinism, while refusing to do so risks rendering such constructs theoretically ineffective.

In response to this tension, many contemporary approaches avoid fully committing to either position. 
Rather than attributing direct causal efficacy to higher-level constructs, they treat such constructs in terms of descriptive, statistical, or informational relations (\cite{hoel, tononi2008, rosas}). 
While this strategy may avoid direct conflict with physicalist assumptions, the existence of such a dilemma does not remove the requirement that theoretical constructs play an explanatory role within a scientific theory. 

The challenge, therefore, is often presented as a choice between physical determinism and mental causation. 
However, this apparent dilemma relies on treating physical determination as equivalent to causation. Physical theories describe how system states evolve, but they do not themselves define causal relations (\cite{pearl, blanchard, woodward}). Consequently, the notion of physical causal closure does not follow directly from physics, but from the additional assumption that physical determination exhausts causation (\cite{ohmura_cs}).

If physical determination and causation are conceptually distinct, then physical determinism and mental causation need not be regarded as competing explanations. Rather, they may be represented within different but compatible descriptive frameworks. The task is therefore not to choose between them, but to formulate a scientific framework in which both can coexist without contradiction.

In the following section, we develop such a framework by distinguishing between physical and causal descriptions as complementary representational systems.

\section{The Causal Stance: A Framework for Defining Causal Relations}

The discussion in the previous section suggests that the difficulty in introducing causation into theories of mind does not arise from the notion of causation itself, but from how it is related to physical description. 
To address this issue, we begin by clarifying the relationship between different forms of scientific representation.

The perspective that causal and physical descriptions constitute distinct forms of scientific representation is consistent with a broadly pluralistic view of scientific explanation. 
In philosophy of science, it is widely recognized that different scientific domains employ distinct representational frameworks (\cite{giere, fraassen}), each with its own descriptive vocabulary and explanatory aims. 
For example, physical, biological, and psychological theories often characterize the same systems in terms of different variables and structures, without requiring a single unified description

In scientific practice, such coexistence of multiple representational frameworks is not only accepted but indispensable. 
Higher-level descriptions are routinely introduced when they provide explanatory or predictive advantages, even when they are not reducible to lower-level formulations in a straightforward manner. 
Accordingly, treating causal and physical descriptions as distinct frameworks does not introduce a new ontological commitment, but reflects a standard methodological feature of scientific theory construction.

Within this perspective, we distinguish between physical and causal descriptions. 
A central feature of causal relations is their asymmetry. 
In contemporary causal modeling frameworks, such as those developed by \cite{pearl}, causation is represented in terms of directed relations between variables, in which causes and effects are not interchangeable. 
This asymmetry plays a crucial role in determining how interventions on a system propagate and affect its behavior.

By contrast, physical descriptions specify how system states evolve according to dynamical laws. 
They characterize lawful state transitions, but do not, in themselves, distinguish between causes and effects. 
In this sense, physical determination and causal relations are conceptually distinct: the former concerns state evolution, whereas the latter introduces an additional asymmetric structure that is not specified by physical laws alone.

Causal descriptions provide distinct advantages when the aim is to represent the internal organization of a system. 
First, they make explicit the asymmetry between causes and effects, which is essential for identifying directionality in the generation of system behavior. 
Second, causal models do not admit transitions between identical states as causal relations. 
Because causation is defined in terms of asymmetric dependencies, it requires differences that make such asymmetry possible.

These features allow causal descriptions to represent internal structure with greater transparency than purely dynamical models. 
In physical descriptions, different internal organizations may give rise to identical state transitions, making it difficult to distinguish them at the level of system behavior. 
By contrast, causal descriptions require the explicit specification of relations that generate these transitions, thereby making differences in internal structure more directly observable within the model.

Importantly, while it is often possible to obtain a physical description from a causal one by disregarding asymmetry, the reverse transformation is not straightforward. 
Recovering causal structure from physical state evolution requires identifying asymmetric relations that are not explicitly encoded in the physical description. 
Since no general method exists for reconstructing such asymmetry (\cite{pearl}), causal descriptions provide a more direct framework for representing the internal mechanisms underlying system behavior.

\begin{figure}[tb]
\begin{center}
\includegraphics[width=8cm]{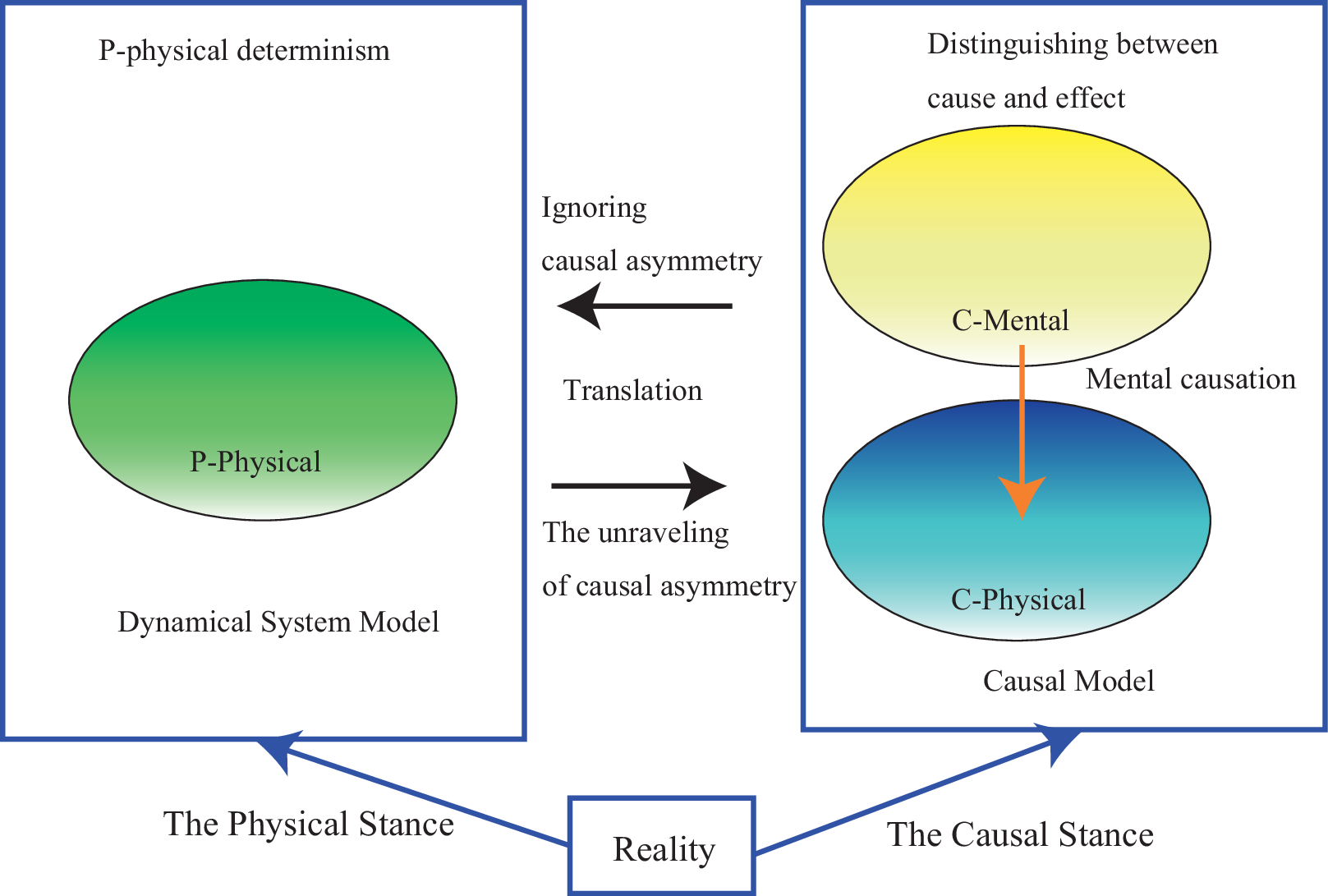}
\caption{
The Physical Stance and the Causal Stance. In the Physical Stance, P-physical states are determined by P-physical states alone and the causal asymmetry between cause and effect cannot be described. In the Causal Stance, identifying this asymmetry allows us to create a structured framework that distinguishes between C-mental and C-physical. C-physical states are not determined by C-physical states alone because of mental causation. When asymmetry is ignored, the causal model is translated into a physical model and must satisfy P-physical determinism. Satisfying P-physical determinism prohibits the introduction of nonphysical forces or entities. This framework is a descriptive system that does not commit to any particular ontology.
}
\label{fig:causal_stance}
\end{center}
\end{figure}

On this basis, we distinguish between two complementary perspectives: the ``Physical Stance'' and the ``Causal Stance'' (\cite{ohmura_cs}). 
The Physical Stance represents systems in terms of states and dynamical laws governing their evolution, while the Causal Stance represents systems in terms of variables and asymmetric relations that capture how interventions affect behavior.

Within this framework, we aim to define mental states and mental causation in terms of the Causal Stance, thereby constructing a scientific framework capable of accounting for the causal role of consciousness. 
In doing so, we address the dilemma outlined in the previous section. 
We therefore proceed by specifying the conditions that such a framework must satisfy.

To develop a descriptive system that can account for the mind and mental causation, the concept of ``physical'' in the Physical Stance must have a different meaning from that in the Causal Stance (\cite{ohmura_cs}).
This is because, in the Physical Stance, changes in physical states are determined solely by physical states and therefore cannot be simultaneously influenced by nonphysical mental states. To distinguish the meaning of ``physical'' in each context, we define the physical state in the Physical Stance as P-physical and that in the Causal Stance as C-physical. We then propose the following conditions to construct a scientific framework for defining mental causation (Fig. \ref{fig:causal_stance}):  

\begin{itemize}
\item Translation: a model in the Causal Stance can be translated into a model in the Physical Stance.
\item P-physical determinism: in the Physical Stance, changes in P-physical states are determined by P-physical states alone.
\item C-mental causation: in the Causal Stance, changes in C-physical states can be determined by both C-mental and C-physical states.
\item Intrinsic causes: in the Causal Stance, mental causation relies on intrinsic causes.
\end{itemize}

\section{Intrinsic Causes}

It is not easy to define intrinsic causes. 
Like volition (\cite{haggard2018}), which is explained in terms of intrinsic causes, the concept may be easier to characterize through negative statements. Volition excludes actions caused by external stimuli as well as those driven by physiological needs within the system. Thus, it is insufficient to assume that intrinsic causes lie within the system itself because the internal triggering of an event in the internal organs (i.e., guts) does not capture what we mean by volition. We therefore define ``intrinsic cause'' by the following three conditions:

\begin{enumerate}
    \item System A, where the intrinsic cause originates, and System B, where the effect occurs, are inseparable.
    \item System A and System B are inseparable because they share the same physical entities.
    \item If the shared entities are removed, the asymmetry between cause and effect is lost.
%    \item The intervention of intrinsic causes does not immediately change the state of shared physical entities.
\end{enumerate}

System-level causality does not exist when a system is broken down into its constituent elements. Condition 1 enables us to treat Systems A and B as a single, integrated system. This allows causes to originate within the system and to produce effects within the same system. Condition 1 is similar to the postulate of ``integration'' in Integrated Information Theory (IIT) (\cite{tononi2008, tononi2015}). 
However, the definition of separability differs. While IIT determines separability based on changes in integrated information, we use shared physical entities as our criterion. This is not merely physical contact, such as contact between internal organs and nerves, but requires the sharing of objects such as cells. 
This criterion is determined by the physical structure. 
Condition 3 is necessary to eliminate trivial cases, such as cases in which shared entities do not contribute to the causal relationship. 
Although further discussion is needed to determine whether our definition is sufficiently robust to exclude negative examples, it has the promising characteristic of being observer-independent and well-defined in the Causal Stance.

To realize our goal, we must identify intrinsic causes. To do so, we examine the conditions under which a model of causal mechanisms within a system generates intrinsic causes. As a candidate mechanism, we propose a model of whole-to-parts causation (\cite{ohmura_wp}). 
Notably, we do not claim that whole-to-parts causation is the only way to explain intrinsic causes. Rather, we consider the above conditions to be necessary requirements for a scientific framework that can account for mental causation.

There is a long history of considering mental phenomena as macrophenomena (\cite{broad, sperry1991}),
and many theories of consciousness seem to regard consciousness as an integrated macro-level phenomenon (\cite{baars2021, crick, freeman,kuhn, tononi1998}). 
However, if a coarse-grained state is determined bottom-up by lower substrate-level neural states (\cite{crick}),
then coarse-grained states cannot have causal effects on the substrate-level (\cite{kim1998}).
Kim’s exclusion argument is often interpreted as denying causality between hierarchical structures (\cite{bontly}).
However, in Kim’s argument, the macro-level state is not defined as a hierarchical level comprising
multiple macro-entities; therefore, it is inappropriate to interpret Kim’s argument as a denial of whole-to-parts causation from a higher to a lower level.

Single-level physical or causal models are commonly used to describe phenomena across different hierarchical levels, such as those examined in sociology, psychology, neuroscience, chemistry, and quantum mechanics (\cite{gillett, laughlin}).
However, such models do not account for whole-to-parts causation. Because Kim’s argument does not rule out whole-to-parts causation, we assume that the underlying causal mechanisms should not contradict physical determinism. By examining the conditions required to realize whole-to-parts causation in physical systems, we derive a model of internal mechanisms for a system with distinct dynamics across two hierarchical levels.

Specifically, we propose a dual-laws model (DLM) designed to describe independent dynamics and their causal influences at different hierarchical levels, namely the coarse-grained and substrate levels. By introducing the causal assignment operator (\cite{pearl}), the DLM differs from single-level dynamical system models.

In the DLM, we construct a descriptive framework that defines the dynamics of the higher coarse-grained level as C-mental laws, the dynamics of the lower substrate level as C-physical laws, and intrinsic whole-to-parts causation as C-mental causation. When translated into the physical model, the DLM satisfies P-physical determinism. Using the DLM framework, we examine the generation of consciousness and the causal efficacy of consciousness.

In this framework, the asymmetry between cause and effect is treated as a primitive concept of causal analysis, in line with the standard causal science developed by \cite{pearl}.
Physical models do not encode causal direction, and the task of grounding the asymmetry of causation is outside the scope of this work. Instead, we focus on how asymmetric causal roles are implemented within a hierarchical structure and articulated explicitly by causal notation.

\section{Dual-Laws Model}
The DLM describes the causal constraint from the coarse-grained level to the substrate-level. 
It differs from interaction models in the Physical Stance involving two distinct dynamics at the same hierarchical level. Uniquely, the DLM possesses a self-referential feedback control mechanism through which causation is transmitted from a higher level to a lower level. 
Because the coarse-grained level (where causes arise) and the substrate level (where effects arise) share the same physical entities, this mechanism allows the introduction of a hierarchical structure, intrinsic causes, and causal asymmetry. In contrast, physical entities are not shared by two dynamics at the same hierarchical level. As a result, the two dynamical systems are separable.

Causal models and dynamical system models differ in their descriptive approaches. Causal models explicitly distinguish the asymmetry between cause and effect, whereas physical models do not. The DLM is therefore a causal model defined within the Causal Stance.

The distinction between cause and effect is not introduced solely to differentiate the DLM from dynamical system models. Elucidating causal structures is a critical task in applied science and engineering, and causal models should be clearly distinguished from dynamical system models. Furthermore, introducing asymmetry in causal inference changes the available statistical methods (\cite{pearl}).
In traditional philosophical discussions of the mind, the distinction between causality and physics has often been unclear
(\cite{kim1998, searle}).
We believe that differences in the underlying models give rise to differences in scientific methodology.

\subsection{Causal Efficacy of Consciousness}
Theories of mental causation, such as the causation of consciousness, are often argued to be flawed because they do not satisfy physical causal closure. 
However, we disagree with this conclusion. According to the currently dominant theory of causation (\cite{pearl, woodward}), the language of physics does not contain asymmetries based on manipulability; therefore, physical causal closure cannot be defined in this framework (\cite{ohmura_cs}).
What physics must uphold is physical determinism, which proposes that changes in physical states are determined solely by physical states. As a result, nonphysical forces cannot be incorporated into this model.

The science of consciousness models the brain as being composed of physical entities that can be explained by physical forces alone. Although causation must satisfy the laws of physics, the laws of physics do not describe manipulability or asymmetry.

Our aim is to create an internal mechanism model of the brain that explains both the causes of consciousness and the causal efficacy of consciousness, with respect to its functional aspects, from the Causal Stance. To emphasize the asymmetry between cause and effect, we identify an intrinsic cause that can be manipulated within the system. 
We then introduce an assignment operator to construct a whole-to-parts causal model that is distinct from a single-level dynamical system model. 

Mental phenomena, including consciousness, are thought to systematically covary with changes in substrate-level neural states, such that no change in the former occurs without a change in the latter. 
In philosophy, this dependency is called a supervenience relationship (\cite{kim1998}). 
For example, the sensation of pain is considered to arise from activity in the peripheral nerves or cerebrum. Changes in supervenient phenomena do not occur without changes in subvenient phenomena. However, this does not mean that supervenient phenomena are caused by subvenient phenomena. Mereological relationships, such as those between whole and parts, also fall under supervenience relationships, although they are more about cooccurrence than causation.

Kim's exclusion argument is often interpreted as denying whole-to-parts causation. However, his argument does not concern causality between levels; rather, it considers the macro-level as a single supervenient entity. Because our model assumes a hierarchy composed of multiple supervenient entities, it falls outside the scope of Kim’s exclusion argument and we do not challenge it. To support the view that intrinsic causes within a system are involved in the generation of consciousness, we examine physical systems capable of realizing whole-to-parts causation.

First, it is important to distinguish between predictability and causality (\cite{sanchez}).
For example, the current position of an object is a good predictor of its position in the immediate future; however, this relationship is not causal. Causality is a relationship that arises between different objects, not between the same physical state at different times. 
In nonlinear dynamical systems, coarse-grained states often exhibit behaviors that appear autonomous or to exhibit whole-to-parts causation (\cite{rosas, seth2010}). 
However, from the manipulability perspective, this apparent whole-to-parts causation measured by predictability is difficult to consider as true causality. 
In true causality, it is possible to identify a cause that does not influence an effect without a causal transmission mechanism 
(\cite{salmon, pearl}).
To define whole-to-parts causation, we must address causes at the supervenient-level (\cite{craver}).
We cannot use the causal assignment operator to distinguish between cause and effect without identifying a supervenient-level cause. Because a single supervenience relationship does not allow us to define a cause at the supervenient level, we assume a hierarchy consisting of multiple supervenience relationships.

To address the functional aspects of consciousness, we make the following assumptions in the Causal Stance:

\begin{itemize}
\item All mental entities, including consciousness, supervene on C-physical states.
\item Whole-to-parts causation, defined as causal efficacy from the supervenient level to the corresponding subvenient states, is involved in the generative mechanism of consciousness.
\item Consciousness is a supervenient, coarse-grained level entity, and the C-physical correlates of consciousness are described by the states of the corresponding subvenient entities.
\item Generated consciousness has causal efficacy indirectly through the same whole-to-parts causation mechanisms.
\item Intrinsic whole-to-parts causation is a model of mental causation.
\end{itemize}
Note that these assumptions do not use the language of physics; here, ``physical'' refers to C-Physical. 
Furthermore, in the Physical Stance, we must maintain physical determinism, which holds that changes in P-physical are determined solely by P-physical states. The causal model can be transformed into a physical model by ignoring the asymmetry based on manipulability to confirm this.

In our discussion of the causality of consciousness and the mind, readers may mistakenly believe that we are introducing nonphysical forces. Rather than introducing nonphysical forces, our goal is to develop a model of the causal efficacy of consciousness in the Causal Stance. Our model introduces the distinction between cause and effect using the causal assignment operator, not a nonphysical force. This distinction does not exist in the language of physics.

\subsection{Self-referential Feedback Control Mechanism}
We assume that whether a system possesses consciousness should be determined by its internal causal mechanisms, rather than by the input–output relationships of the system or its subsystems (\cite{fahrenfort2021}). To clarify these intrinsic causes, we focus on whole-to-parts causation.

To define whole-to-parts causation, we must define a cause at the supervenient level and a causal transmission mechanism from the supervenient level to the subvenient level. Self-referential feedback control mechanisms are necessary for transmitting causation from the supervenient level to subvenient level.

The mereological relationship between the whole and its parts is a type of supervenience relationship because changes in the whole cannot occur without changes in the states of its parts. Since the whole and its parts share the same physical entities, causation from the whole to its parts can be characterized as intrinsic. 
Our model assumes causality across hierarchical levels based on multiple supervenience relationships. In this case, causation from the whole level to the parts level can be characterized as intrinsic because the two levels share the same physical entities. Our model includes physical entities in addition to those shared across levels. However, when the physical entities shared between levels are removed, causality disappears. This means that the whole-to-parts causation model can explain intrinsic causes.

The relationship between cell groups, such as organs, and their constituent individual cells is a well-known whole-parts relationship in biological systems. In the brain, these cell groups form neural networks. Neural networks can be modeled as functions. We assume that supervenient entities are mathematical functions and that the supervenient level comprises multiple supervenient functions or neural networks. The corresponding subvenient entities are the states of neurons and synapses within the supervenient functions.

An equation can be defined by selecting and ordering multiple supervenient functions from a set of supervenient functions at the supervenient level. We define the self-referential feedback control mechanism as negative feedback control that regulates the states of neurons and synapses at the subvenient level to satisfy the equations composed of the selected supervenient functions. A unique feature of this model is that only the subvenient states corresponding to the selected supervenient functions that constitute the equation are affected by the self-referential feedback control mechanism. Because the coarse-grained supervenient functions and the corresponding subvenient states share the same physical entities, it is possible to introduce a hierarchical structure.

Notably, structural hierarchies are defined by the relationship between the whole and its parts, not by hierarchies in information processing, such as the primary and secondary visual cortices. Because cause and effect arise from the same physical entities in this relationship, the model is capable of explaining intrinsic causes. The following subsection presents this formulation in detail.

\subsection{Formalization}
The formulation presented here is not intended as a perfect model, but as a way to characterize our ideas.

Let $x^i \in \mathbb{R}^{n_i}$ denote the subvenient state variables (e.g., the states of neurons and synapses) for an index set $i \in I \subset \mathbb{N}$. 
These variables define the corresponding supervenient function
whose domain is direct product of $k_i \in \mathbb{N}$ sets $$\underbrace{ \mathbb{R}^m \times \mathbb{R}^m \times \dots \times \mathbb{R}^m}_{k_i}$$ and codomain is $\mathbb{R}^m$ as follows: $X^i: (\mathbb{R}^m)^{k_i} \to \mathbb{R}^m$. 
Let $\mathcal{F}_{k_i}$ denote the space of functions $f: (\mathbb{R}^m)^{k_i} \to \mathbb{R}^m$. 
For each $i$, let $b_i:\mathbb{R}^{n_i} \to \mathcal{F}_{k_i}$ be a coarse-graining map such that 
$X^i = b_i (x^i)$. We only assume that $b_i$ is deterministic and that $X^i$ does not vary without a corresponding variation in $x^i$. 
We are considering the possibility that the number of supervenient functions and the cardinality of the index set may vary over time, but for the sake of simplicity, we have not included this in our formulation.

Let ${X^i}_{i \in I}$ denote a family of coarse-grained (supervenient) functions. 
Equations are defined by selecting and ordering multiple
supervenient functions from this family.

Consider an index sequence $c = [i_0, i_1, ...] \in C$, where $i_k \in I$. 
Each index corresponds one-to-one with an element ${X^i}_{i \in I}$. 
Let $\mathrm{Expr}(\mathcal{X})$ denote the set of well-formed expressions generated from function symbol $X^i$ using composition. 
Each index sequence $c$ defines an expression $\mathrm{E}(c) \in \mathrm{Expr}(\mathcal{X})$, which evaluates to a function $e_c: \mathbb{R}^m \to \mathbb{R}^m$.  
This construction is analogous to functional programming,
where programs specify how functions are composed
rather than passing functions as numerical arguments.

The error function can be represented by an algebraic formula composed of the supervenient functions, with the order of operations determined by the index sequence. We assume that the equation is satisfied when the output of this error function is zero. 
Array of multiple index sequences $\mathbf{c} = [c_0, c_1, ...]$ is a discrete supervenient-level state. 

Next, we introduce a causal relationship between the feedback error and the index sequences.
The feedback error $err$ can be calculated by error function $e_c$ and input $d \in \mathbb{R}^m$ like $err = e_c(d)$. 
Using Pearl's causal operator ``$:=$'' (an assignment operator) (\cite{pearl}), we describe the feedback error  by $err:=e_c(d)$. 
The symbol $:=$ means ``is determined by,'' and Pearl uses $=$ to express this meaning. 
In this formulation, $e_c$ and $d$ are the cause and $err$ is the effect. 
Similarly, we can define multiple error functions for the array of index sequences $\mathbf{c}$. 

We believe that considering changes in the error function as causes at the supervenient level is crucial for defining whole-to-parts causation. Our model therefore incorporates the language of causation to determine the feedback error from the error function. 
According to Pearl’s theory, distinguishing cause from effect in causal graphs requires an asymmetric operator. Without identifying this asymmetry, this operator cannot be introduced.

The causal operator ``:='' cannot be used unconditionally (\cite{bochman}).
For example, in the laws of physics, the relationship between force $f$, acceleration $a$, and mass $m$ can be expressed as $a := f/m$ because force can change acceleration.
However, it is difficult to change force using acceleration, so we cannot write $f/m := a$. 
In other words, causal relationships require a physical structure that goes beyond the laws of physics. Because we defined the supervenience relation as non-causal, we cannot write $X^i := b_i (x^i)$. 
Furthermore, since a change in $X^i$ cannot occur without a change in $x^i$, a relation such as $x^i := F(X^i)$ is impossible.

Our formulation includes two innovations. First, although supervenient entities have historically been treated as vector quantities (\cite{flack,hoel, marshall2026, rosas}),  we formulate them as functions. 
We believe that the traditional approach of modeling coarse-grained states as vectors is influenced by statistical mechanics. 
In contrast, it is common to model neural networks as functions in machine learning and artificial intelligence. This allows us to perform composition operations on supervenient functions and gives meaning to the order of operations. Because vectors are commutative, the order of operations does not matter.

For example, an algebraic constraint such as commutativity ($X^i X^j = X^j X^i$) can be defined between two supervenient functions $X^i$ and $X^j$ (\cite{nishitsunoi}). 
Alternatively, group axioms may be applied to a set of supervenient functions and represented by an equation. In the self-referential feedback control mechanism, the subvenient-level states are controlled so that they satisfy the equations determined by the index sequences. As a result, the equation can be regarded as equivalent to the error function.

Second, we note that the index does not change in response to changes in the corresponding subvenient states. Therefore, the index sequence can have dynamics that are independent of the subvenient state. By considering changes in the error function as causes in a whole-to-parts causal relationship, the supervenient functions that constitute the error function and the subvenient entities changed by the feedback error are physically identical. 

In the self-referential feedback control mechanism, we define changes in the error function as a supervenient-level cause. This supervenient-level cause is the structure-level change produced by the selection and ordering of indexes; it does not directly change the subvenient states. To change the error functions, the changes in the index sequences must have distinct dynamics from the substrate-level dynamics. Importantly, the selection and ordering of indexes are not determined by the underlying supervenience relations. This independence is essential for treating supervenient-level causes as genuine causes in the interventionist sense. It also allows the model to represent hierarchical causation without collapsing supervenient-level causes into mere redescriptions of subvenient dynamics or overdetermination. To transmit causal influence, a causal transmission mechanism is necessary. In our model, this mechanism is self-referential feedback control that acts to satisfy the equations determined by the index sequences. Negative feedback control can be achieved using the gradient descent method. Furthermore, a system with a self-referential feedback control mechanism requires feedback connections, like in recurrent processing theory 
(\cite{lamme}), although feedback connections alone are insufficient for negative feedback control.

\subsection{Dual-Laws Model}
\begin{figure}[tb]
\begin{center}
\includegraphics[width=8cm]{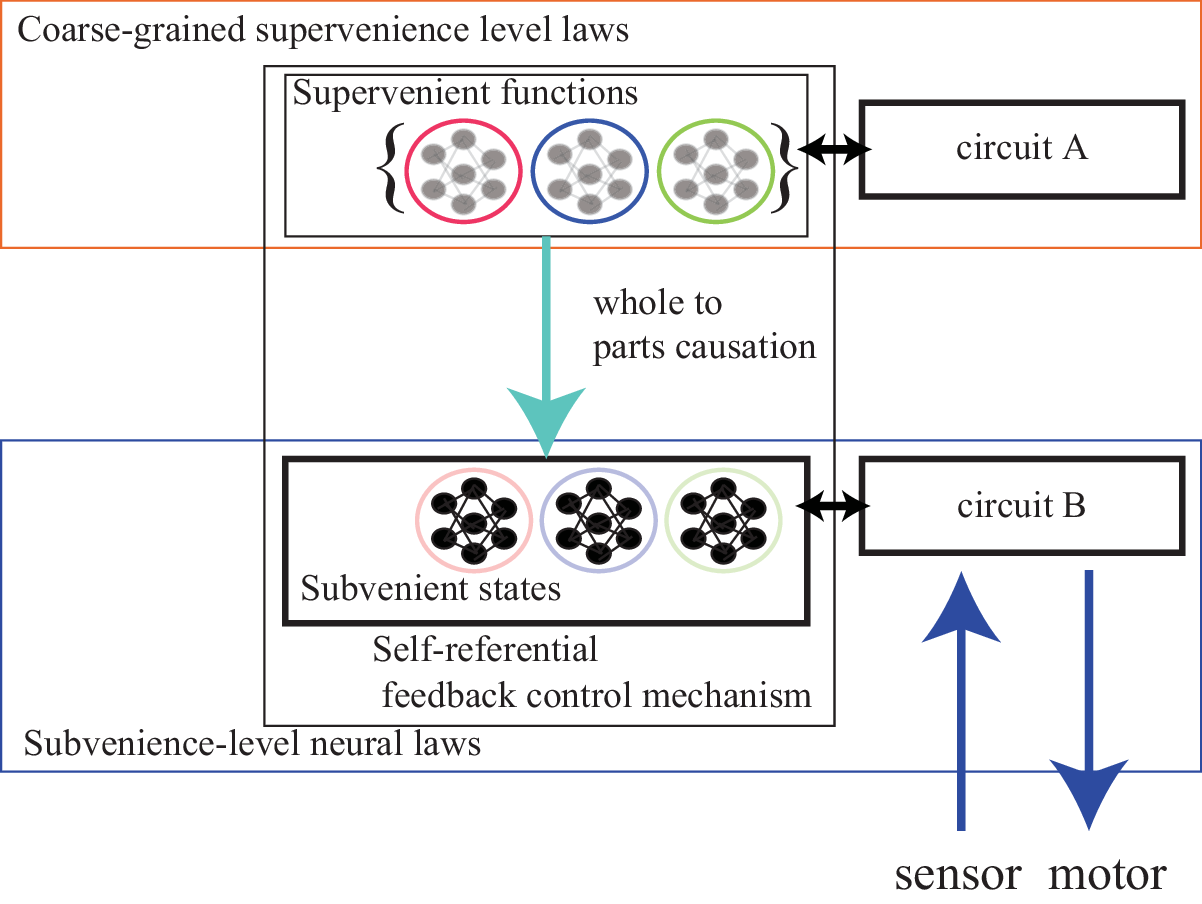}
\caption{
Dual-Laws Model: The supervenient level consists of multiple supervenient functions. The feedback error is the error involved in satisfying the equation defined by combining these supervenient functions. These equations vary according to the index sequences. The dynamical laws at the supervenient level modify and select these discrete index sequences. At the same time, the feedback error can be adjusted by subvenient states (e.g., neurons, synapses, and other components) that constitute supervenient functions through a self-referential feedback control mechanism. Thus, the feedback error is influenced by the both the index sequences at the supervenient level and by the subvenient states, making the Dual-Laws Model possible. Rectangles with bold outlines represent distinct physical entities. In the self-referential feedback control mechanism, supervenient functions and their corresponding substrates share the same physical entities.
}
\label{fig:dual_process}
\end{center}
\end{figure}

Our proposed system model has distinct dynamical systems at the supervenient and subvenient levels (Fig. \ref{fig:dual_process}). 
The system consists of three components: a self-referential feedback control mechanism, supervenient level neural circuits A that modify the index sequences that define the error functions, and other subvenient-level neural circuits B.

Let $t, T \in \mathbb{N}$ denote discrete time steps. 
Let $p$ denote a (possibly nonlinear) state-transition function describing the lower-level dynamics. 
The dynamics of subvenient states can then be described as follows: $ x^i_{t+1}, err_{t+1}, x^b_{t+1} = p (x^i_t, err_t, x^b_t)$. 
We assume that $x^b$ is a state of circuit B, which is involved in the sensory, motor, and life-supporting functions. Therefore, through interaction with neural circuit B, the subvenient states can receive sensory motor information about the body and environment.  

Let $P$ denote a state-transition function governing the dynamics of index sequences at the supervenient level. 
The dynamics of index sequences can be described as follows: $\mathbf{c}_{T+1}, x^a_{T+1} = P (\mathbf{c}_{T}, x^a_T)$. 
We assume that $x^a$ is a state of circuit A.  
We also assume that $T > t$, because the dynamics of equations to calculate feedback error are slow enough 
to allow the self-referential feedback control to converge.
Although an interaction between neural circuits A and B is theoretically possible, it is not considered here. The key point is that the feedback error is determined by the index sequences and can be expressed as $err := e_c(d)$. 

If we replace the causal operator $:=$ with $=$, 
the DLM can be expressed by a single-layer dynamical system model in the Physical Stance: $x_{t+1} = f_{c_{T=t}}(x_t)$, which represents timevariant dynamics. 
The system's internal states are structured, and some of the transition functions change.
Here, $x_t$ consists of all subvenient states and $x^a$. The dynamics of $c$ are determined by the physical states representing $c$ and $x^b$. 
Importantly, this means that the DLM satisfies P-physical determinism. 
If we do not introduce the asymmetry between cause and effect, the DLM cannot be distinguished from a single-level dynamical system model.
To adequately express whole-to-parts causality, it is crucial to distinguish between cause and effect in the model. The key point is that, because physical models do not account for causal asymmetry, causal relationships between hierarchical levels are hidden within the physical model.

Thus, we examine theories of consciousness using the DLM that we defined in terms of the Causal Stance, which is broader than the Physical Stance. Although the Causal Stance may not be necessary to simply describe consciousness, we believe it is necessary to clarify the causes of consciousness. If consciousness arises from within the system, the model should include the asymmetry between intrinsic causes and effects.

\subsection{DLM for a theory of consciousness}
Because consciousness is a supervenient-level phenomenon, we have modeled it in terms of supervenient functions selected by the index sequences that satisfy certain conditions. 
We consider causal efficacy from the supervenient level to subvenient level, mediated by the self-referential feedback control mechanism, to be a necessary condition for the generation of consciousness. We therefore expect that consciousness does not arise in systems that lack a whole-to-parts causation mechanism. Furthermore, the C-physical correlates of consciousness exist only within the subvenient entities selected by the index sequences. The selection of supervenient functions via the index sequence can be regarded as an attentional mechanism. We believe that the properties of this attentional mechanism account for key features of consciousness, such as integration, compositionality, and exclusivity. Because index sequence selection serves as an attentional mechanism, bottom-up influences are also likely to play a role in this selection. However, we do not address this possibility.

Because the generation of consciousness is assumed to require a self-referential feedback control mechanism, this suggests that consciousness does not emerge when this feedback control stops. In the DLM, the internal mechanism can control the on/off state of consciousness. Thus, we propose that the presence or absence of consciousness is inherently binary.

Although we consider the self-referential feedback control mechanism a necessary condition for consciousness, whole-to-parts causation should not be equated with the causal efficacy of consciousness itself. Rather, we propose that generated consciousness indirectly influences the C-substrate level by affecting the index sequences. In other words, generated consciousness changes attention, which subsequently affects subvenient states through the self-referential feedback control mechanism.

\section{Relationship with Existing Theories}
To characterize our theory, we now compare it with existing theories. 
The goal of this comparison is to distinguish our theory from existing ones, not to evaluate our theory’s superiority. 
The DLM is an abstract mechanism and should be distinguished from any specific implementation in the brain. 
We do not consider the DLM to be mutually exclusive with current neuroscience-based theories of consciousness. 
Rather, we view it as a tool for formulating theories from the Causal Stance.

Approaches that infer the generative mechanism of consciousness from neural correlates (\cite{white}) are difficult to verify because they fall into the ``triviality problem'' (\cite{kleiner}).
Our methodology infers the generative mechanism of consciousness based on the conditions necessary for it to possess causal efficacy. 
Rather than introducing a theory within an existing framework, we reformulate the representational basis itself so that mental causation can be incorporated as a fundamental component of scientific models.

\subsection{Functionalism}
According to \cite{block1996}, ``Functionalism says that mental states are constituted by their causal relations to one another and to sensory inputs and behavioral outputs.''
Similarly, the Stanford Encyclopedia of Philosophy (\cite{janet}), states that ``functionalist theories take the identity of a mental state to be determined by its causal relations to sensory stimulations, other mental states, and behavior.'' 
Despite functionalism’s emphasis on causality, an explicit discussion of the asymmetry between cause and effect has not incorporated. Functionalism is often characterized by the relationship between inputs and outputs, although \cite{block, searle1980} consider this to be a problem

Our model does not characterize mental causation or consciousness solely in terms of the input–output relationships of systems and subsystems. When a model of whole-to-parts causation is introduced into a system, the system’s inputs and outputs are not essential to the formulation. The problem with characterizing a system based on similarities in its input–output relationships is that this obscures the system’s internal mechanisms. 
If the goal is to emphasize the intrinsic nature of a generative mechanism of consciousness, it is necessary to distinguish the system’s internal mechanisms. A causal mechanism model that focuses on the asymmetry between cause and effect is well-suited to this purpose. Our position is that the mind and consciousness should be characterized not by the similarity of a system’s input–output relationships, but by the causal mechanisms within the system itself.

\subsection{Global Workspace Theory}
Global Workspace Theory (GWT) links consciousness to broadcasting functions (\cite{baars1988,  baars2021,  dehanene}).
It is relevant because it has been used to emphasize the functional role of consciousness (\cite{baars1988}). 
We consider theories of consciousness that ignore functions to be scientifically inadequate (\cite{cohen, herzog}). 
However, the idea that function alone generates consciousness faces the typical problems of functionalism (\cite{block, chalmers1995}). 
GWT posits that parallel distributed processing is unconscious, whereas serial, integrated, and coherent information broadcasted to distributed systems is conscious
(\cite{baars1988, baars2021}). 

In contrast to what \cite{baars1988} proposed, we argue that a function like broadcasting is not a necessary condition for generating consciousness. We claim that consciousness does not arise bottom-up from substrate-level dynamics, but through whole-to-parts causation. While the discrete nature of index sequences and the selection mechanism may relate to the seriality, integration, and coherence of consciousness, we believe that the generative mechanism of consciousness and the causal efficacy of consciousness are distinct problems. In our view, broadcasting is more closely related to the causal efficacy of consciousness than to the generative mechanism of consciousness. Moreover, we do not assume that localized brain regions or pathways like the global workspace are necessary to generate consciousness. Instead, we hypothesize that a self-referential feedback control mechanism is necessary for the generation of consciousness.

\subsection{Freeman's Circular Causation}
In the present study, we describe the mechanism by which the supervenient level causally constrains its corresponding subvenient states as a self-referential feedback control mechanism. This differs from Freeman’s Circular Causation
(\cite{freeman}), in which elements at the same level interact to form interaction loops in the Physical Stance.
In such complex systems, global states emerge from micro-level interactions and then have downward organizational effects on individual neurons.  
In contrast, the coarse-grained dynamical laws we propose are not obtained through self-organization at the substrate level. \cite{bedau} notes that macrophenomena arising from self-organization cannot explain ``real'' whole-to-parts causation. 
The key question is thus whether the asymmetry between cause and effect emerges from the interactions within physical systems or if the language of causation should be added to the language of physics.
We aim to achieve the latter by developing a whole-to-parts causation model using the Causal Stance, rather than a dynamical system model using the Physical Stance.

\subsection{Integrated Information Theory}
IIT defines consciousness through five essential axioms (properties) of phenomenal experience: intrinsicality, composition, information, integration, and exclusion
(\cite{tononi2008, tononi2015}).
Although IIT also emphasizes intrinsicality, the meaning differs from our notion of intrinsic cause. In early versions of IIT (\cite{tononi2008}), intrinsicality was mainly associated with observer-independence, whereas recent formulations (\cite{albantakis}) characterize it in terms of a system's intrinsic cause-effect power upon itself. By contrast, our notion of intrinsic cause concerns the origin of causation itself, namely whether the cause is generated within the system rather than imposed externally.

Although \cite{hoel} addressed the relationship between IIT and macrocausation, the macro-states they describe are different from ours. 
They assume that macro-states are coarse-grained states represented by vectors. In contrast, in the DLM, the coarse-grained level states are index sequences that define error functions composed of supervenient functions. Because these coarse-grained level states are not supervenient entities, independent supervenient-level causes can be defined. To define the causal mechanism, we identified both a cause and a causal transmission mechanism within the hierarchical structure.

Both IIT and DLM place a strong emphasis on causal structure, but their definitions of causality differ significantly. IIT characterizes consciousness in terms of an intrinsic cause-effect structure derived from informational constraints among system states and evaluates causation through cause-effect repertoires specified by system dynamics. This notion of causation differs from interventionist accounts such as those of \cite{pearl} and \cite{woodward}. By contrast, our framework treats causal asymmetry and causal transmission mechanisms as primitive features that cannot, in general, be reconstructed from state-transition descriptions alone.

The major difference between IIT and our approach is that IIT assumes integration through a bottom-up process within the recurrent structure of a dynamical system (\cite{albantakis}).
In contrast, we posit a DLM that explicitly incorporates a self-referential feedback control mechanism, which exerts an influence from the supervenient level to the corresponding subvenient states and is formulated using the Causal Stance. We believe that a self-referential feedback control mechanism is necessary to explain the intrinsic cause of the generation of consciousness.

In addition, what we consider to be intrinsic is not consciousness itself, but the cause of its generation. Because consciousness depends on sensory inputs and physiological states, consciousness itself does not appear to be intrinsic. We also do not think that the private nature of conscious experience implies that it is intrinsic. We believe that what is intrinsic lies with the cause of the generation of consciousness.

We also believe that the attention mechanism in our model, which is the selection of discrete index sequences, can explain several features of consciousness, such as integration, composition, and exclusiveness. We do not believe that all properties of consciousness can be explained solely by substrate-level dynamics in the Physical Stance.

%These differences reflect a broader issue discussed in the Introduction, namely that causal relations are often introduced without a clear distinction from physical description.

\subsection{Non-reductive physicalism}
Non-reductive physicalism holds that consciousness is a product of the brain, whereas mental states and properties are non-reductively distinct from physical states and properties (\cite{macdonald}). 
According to \cite{kuhn}, this view resembles property dualism because it treats mental states as ontologically distinct from physical states. The core mechanism of non-reductive physicalism is emergence, which is the idea that higher level emergent properties cannot be predicted even with complete knowledge of the underlying level. Moreover, non-reductive physicalism often assumes that emergent mental properties exert downward causal influence on physical systems. This notion of downward causation is frequently used to explain mental causation (\cite{mayr, sperry1991}) and agent causation (\cite{conner, steward2012, steward2017}).
Non-reductive physicalism assumes that emergent mental properties are ontologically irreducible to physical properties. In contrast, we propose the DLM as a causal model with a self-referential feedback control mechanism. Although the DLM employs the language of causality beyond the language of physics, it does not require ontological emergence. The DLM satisfies P-physical determinism when it is converted to the physical model by ignoring the asymmetry between cause and effect. Thus, our model does not rely on the concept of emergence or nonphysical forces.

\cite{gillett} proposes a non-reductive theory that assumes influence from the macro-level to the micro-level. However, his concept of machretic determination differs from our notion of causal influence. Machretic determination refers to a symmetric relationship of mutual determination between macro- and micro-levels, rather than a directional causal relationship. In contrast, our proposed self-referential feedback control mechanism allows the coarse-grained supervenient level to causally constrain the corresponding subvenient states; we interpret this as a form of causation because we can identify a supervenient-level cause and a causal transmission mechanism separately. In our formulation, we can explicitly use the causal assignment operator because we can identify an asymmetric relationship between cause and effect within the hierarchy.

\subsection{Emergentism}
\cite{sperry1969} considers consciousness as ``the holistic properties of the organism with causal effects,'' which appears to equate coarse-grained intrinsic causation with consciousness. In contrast, we do not equate the causal power from the coarse-grained supervenient level to subvenient states with consciousness itself. We believe that whole-to-parts causation is necessary for generating consciousness, but that how consciousness influences other neural systems is a separate issue. To investigate the causal efficacy of consciousness, it is necessary to clarify how consciousness affects the neural system after it is generated. We assume that consciousness affects the selection of index sequences and subsequently influences the subvenient level through the self-referential feedback control mechanism.

When theories of consciousness emphasize intrinsic causes, they often intersect with the problem of free will or agent causation. \cite{sperry1976} appears to equate the problem of consciousness with that of free will.
Recent discussions have also linked IIT with the problem of agency (\cite{desmond, potter}). 
These connections arise from a shared challenge: coarse-grained intrinsic causality cannot be explicitly explained by the Physical Stance alone. From our perspective, this is because physics does not distinguish between the asymmetry of cause and effect, which reduces the transparency of a system’s internal mechanisms. As a result, physics cannot describe intrinsic causes. The issue thus lies in the descriptive ability of physics, not in the idea that intrinsic causes violate the laws of physics.

At the surface, Scheffel’s Emergent Will resembles our DLM.  \cite{scheffel} considers Emergent Will as an independent law, separate from lower-level physical laws, and proposes it as a new definition of free will. To justify the validity of Emergent Will, Scheffel appeals to the emergence of psychological properties, such as non-reductive physicalism. However, if the DLM is assumed, Scheffel’s proposed concept of free will can be maintained without the need for mysterious causal emergence. Similarly, \cite{list} employs the emergence of intentional agency from lower-level physical phenomena in his discussion of free will.  Our DLM is not a model of emergence, but a whole-to-parts causation model derived from the Causal Stance. To resolve the contradiction between mental causation and physical determinism, we believe it is necessary to distinguish between P-physical and C-physical. C-physical is subject to the causal influence of C-mental through C-mental causation, whereas P-physical is only subject to the influence of P-physical. 

In debates about free will, intuitive resistance to determinism may stem from the assumption that intrinsic causes cannot affect system behavior due to P-physical determinism. In the Causal Stance, the behavior of the C-Physical can be causally constrained by the C-mental. Thus, it is essential to distinguish between P-physical and C-physical. If we assume a DLM in the Causal Stance, we can formulate intrinsic causes as supervenient-level causes and C-mental laws as supervenient-level dynamics.

\subsection{Piaget's developmental theory}
Piaget’s developmental theory is also relevant to our framework. Piaget argued that structural changes in intelligence arise through two mechanisms: assimilation and accommodation (\cite{piaget}). 
Assimilation is the active process of incorporating physical stimuli, such as sensory input, into an organism’s internal structure, whereas accommodation is the process of modifying the internal structure itself.

Piaget presupposed an agent with a mental capacity characterized by these two processes, and appeared to believe that the mechanisms of biological development were not reducible to physical laws alone. 
Although his epistemology was influenced by Kant, he criticized Kant’s static view of \textit{a priori} structures and experimentally demonstrated that logical and mathematical structures evolve during development. Our concept of coarse-grained level dynamics is inspired by Piaget’s idea of structural reconstruction: we interpret assimilation as corresponding to the self-referential feedback control mechanism and accommodation as corresponding to the coarse-grained level dynamics, which changes the algebraic structural constraints. In this way, Piaget’s developmental framework provides a conceptual foundation for understanding how internal structures evolve and interact with the substrate level in our model.

\subsection{Summary}
Existing theories of consciousness do not seem to identify the asymmetry between cause and effect within the system; rather, they posit that this distinction emerges within the dynamical system model. As a result, the language of causality is not explicitly incorporated into existing models. We explicitly introduce the distinction between cause and effect into our whole-to-parts causal mechanism model. In this way, we argue that the Causal Stance is necessary to elucidate the generative mechanism of consciousness.

Introducing multiple dynamics at different levels into a scientific theory may be criticized as a violation of Occam’s razor. However, science already studies different laws independently at each hierarchical level. Assuming multiple dynamical laws is therefore not inherently problematic. In the DLM, the causal influence from the coarse-grained supervenient level to the subvenient level is essential for explaining intrinsic causal power. Thus, the additional complexity of multiple dynamics is justified because it explains phenomena that cannot be captured by a system model lacking a whole-to-parts causation mechanism.

\section{What consciousness can do for neuroscience}
\cite{gomez} suggests that we should ask not what neuroscience can do for consciousness, but how theories of consciousness can contribute to neuroscience.  
We agree with this perspective. To avoid the triviality problem
(\cite{kleiner}), we do not derive our theory directly from neural data. Instead, we propose a causal model intended to constrain and guide future neuroscientific interpretation.

Traditional scientific theories often study laws independently at each hierarchical level. In contrast, we assume that distinct dynamical laws operate at both the supervenient and subvenient levels, and that subvenient-level observations arise from interactions between these levels. Because we formulate our model from the Causal Stance, caution is warranted when applying analytical methods that ignore causal asymmetry. Moreover, verifying causal influence from the supervenient level to the subvenient level requires interventions on supervenient-level causes, which are difficult to realize in actual brain systems. For this reason, a constructive approach may be particularly appropriate.

Causal models are widely used in engineering design, making their connection to constructive approaches a natural one. Reproducing subvenient-level observations under complex whole-to-parts interactions is therefore a nontrivial task that may benefit from constructive modeling.

One option is to implement both levels of dynamics in simulations or robotic systems and study the resulting behavior. Self-referential feedback control mechanisms can be simulated using artificial neural networks, where supervenient functions are realized at a higher descriptive level and subvenient states correspond to models of neurons and synapses that constitute these systems. Such implementations could make it possible to test which models are involved in the functional aspects of consciousness.

Furthermore, if consciousness modulates index sequences, such as those involved in attentional selection, our model could yield predictions about how generated conscious states influence observable subvenient physical dynamics.

Because neither coarse-grained supervenient dynamics nor fine-grained subvenient dynamics can be fully derived from observational data alone, exploratory strategies will likely be required. For example, machine learning methods could be useful for exploring combinations of dual-level dynamics that account for the functional aspects of consciousness.

\section{Conclusion}

A central challenge in the science of consciousness is to explain how internal structures contribute to the generation of system behavior. 
As discussed in this paper, many existing approaches introduce increasingly rich descriptions of internal organization, yet often fail to specify how these structures play a causal role within the system. 
This raises the risk that such constructs become theoretically ineffective, despite being described in causal terms.

We have argued that this difficulty arises in part from a lack of distinction between physical and causal descriptions. 
When causal relations are implicitly identified with physical state evolution, the role of causation in generating behavior remains unclear. 
Relatedly, treating predictive or information-theoretic relevance as indicative of causation obscures the distinction between explanation and prediction.

To address these issues, we introduced a framework based on the explicit separation of two complementary perspectives: the Physical Stance and the Causal Stance. 
Within this framework, causal relations are defined in terms of asymmetric dependencies, independent of physical determination. 
This allows causal descriptions to provide a more constrained and transparent representation of the internal mechanisms that generate system behavior.

Within this framework, our main contribution is the introduction of the Dual-Laws Model (DLM) as a causal framework for consciousness. 
Unlike existing approaches, the DLM represents consciousness in terms of explicit internal causal mechanisms, allowing the presence of consciousness to be determined from the structure of the system itself rather than from observation or prediction. 
In this sense, it provides a transparent, design-level model of consciousness based on intrinsic causal organization.

Importantly, the DLM satisfies physical determinism when expressed in the Physical Stance, while enabling well-defined causal relations in the Causal Stance. 
In this way, the framework reconciles the requirement of physical determinism with the introduction of causally effective higher-level structure.

More generally, our analysis suggests that progress in the science of consciousness requires not only more detailed descriptions of internal structure, but also a clear specification of the representational framework in which causation is defined. 
By explicitly incorporating causal asymmetry into the description of internal mechanisms, the present framework provides a principled basis for modeling the generative role and causal efficacy of consciousness.

\section{Acknowledgments}
The authors thank the anonymous reviewers for their valuable suggestions. This research was supported by JSPS KAKENHI Grant Number JP25K24741, Japan. The funding sources had no role in the preparation of this manuscript or the decision to publish it.

\section*{Author Contributions}
YO: conceptualization, draft writing, and revision; YK: supervision, and funding.

\section*{Availability of data and material}
No new data were created or analyzed in this study.

\section*{Competing interests}
The authors declare no relevant financial or nonfinancial interests.

%UNCOMMENT THE BELOW TWO LINES IN CASE YOU NEED NUMBERED FORMAT.
%\bibliographystyle{oup-plain}
%\bibliography{reference}

%UNCOMMENT THE BELOW TWO LINES IN CASE YOU NEED AUTHOR YEAR FORMAT.
%\bibliographystyle{oup-abbrvnat}
%\bibliography{reference}

\bibliography{main}
\bibliographystyle{apalike}

\end{document}